\documentclass[12pt]{article}

\usepackage{newtxtext,newtxmath}
\usepackage{physics}
\usepackage{dsfont}
\usepackage{xcolor}
\usepackage[normalem]{ulem}

\usepackage{graphicx}

\usepackage[letterpaper,margin=1in]{geometry}

\renewenvironment{abstract}
	{\quotation}
	{\endquotation}

\date{}

\makeatletter
\renewcommand{\fnum@figure}{\textbf{Figure \thefigure}}
\renewcommand{\fnum@table}{\textbf{Table \thetable}}
\makeatother

\usepackage{url}

\def\scititle{
	Universal Protection of Quantum States from Decoherence
}
\title{\bfseries \boldmath \scititle}

\author{
	Francesco~Atzori$^{1,2}$,
	Salvatore~Virzì$^{1}$,
    Francesco~Devecchi$^{3}$,\and
    Domenico~Abbondandolo$^{1,2}$,
    Alessio~Avella$^{1\ast}$,
    Fabrizio~Piacentini$^{1}$,\and
    Marco~Gramegna$^{1}$,
    Ivo~Pietro~Degiovanni$^{1,4}$,
	Marco~Genovese$^{1,4}$\and
	\small$^{1}$ Istituto Nazionale di Ricerca Metrologica, Strada delle Cacce, 91, Torino, 10135, Italy.\and
	\small$^{2}$Politecnico di Torino, Corso Castelfidardo, 39, Torino, 10129, Italy.\and
    \small$^{3}$Università degli Studi di Torino, via P. Giuria 1, Torino, 10125, Italy.\and
    \small$^{4}$Istituto Nazionale di Fisica Nucleare, via P. Giuria 1, Torino, 10125, Italy.\and
	\small$^\ast$Corresponding author. Email: a.avella@inrim.it\and
}

\begin{document}

\maketitle

\begin{abstract}
The fragility of quantum coherence fundamentally limits the scalability of quantum technologies, as unavoidable environmental interactions induce decoherence and rapidly degrade quantum properties.
The Quantum Zeno Effect offers a powerful route to suppress quantum evolution and protect coherence through frequent measurements, irrespective of the underlying dynamics. However, existing implementations require prior knowledge of the quantum state, severely restricting their applicability.
Here we introduce a state- and dynamics-independent protection protocol embedding the system in a larger Hilbert space, temporarily swapping the quantum information from its original degree of freedom to a decoherence-free ancillary one.
We experimentally validate the protocol on a quantum optical platform, demonstrating robust preservation of coherence and purity for arbitrary polarization qubits under decoherence, thereby enabling the universal safeguarding of unknown quantum states.
\end{abstract}

\noindent

Quantum technologies~\cite{Gerard2003} are rapidly advancing across a variety of platforms, including photonics \cite{Kim2017,OBrien2009}, trapped ions \cite{Jeremy2019}, superconducting qubits \cite{Wendin2017}, and many others.
Applications such as quantum computing \cite{Imre2019}, communication \cite{Kumar2024}, metrology \cite{giov2006,gen2021}, imaging \cite{Genovese2016} and sensing \cite{Cappellaro2017,Petrini2020} rely on the ability to prepare, manipulate, and transmit quantum states with high fidelity. However, in all these platforms, the fragility of quantum coherence poses huge technological challenges, severely limiting their widespread diffusion.

In general, the interaction between a quantum system and its surrounding environment leads to the degradation of the purity and the fidelity of quantum states, a phenomenon known as decoherence \cite{Zurek2003}.
This interaction, at the very basis of the macro-objectivation problem \cite{Genovese2010}, can occur through various mechanisms, making it difficult or even impossible to fully characterize and/or eliminate.

As a result, several strategies have been developed to study and mitigate decoherence \cite{Zanardi1997,Viola1998,Lidar1998,Vitali1999,Viola1999,Kofman2000,Kofman2001,Kim2012, Souza2012,Piacentini2017,Virzi2022,Virzi2024}.
Among them, protocols inspired by the Quantum Zeno Effect (QZE) \cite{Sudarshan1977,Wineland1990} leverage Quantum Zeno dynamics \cite{Tasaki2006,Kondo2016,Facchi2008}, wherein repeated observations of the quantum system can inhibit decoherence by freezing its evolution and keeping it in its initial state, or confining it in a subspace of the original Hilbert space.
However, these protocols typically require prior knowledge of the quantum state or the type of system-environment interaction, as the protection must be tailored specifically to the particular case.

Here we propose and experimentally demonstrate a universal protection protocol, named Quantum State Universal Protection (QSUP), that lifts this requirement.
We show that one can transfer the unknown quantum superposition constituting the system initial state to a suited, decoherence-free ancillary degree of freedom (DoF), bringing the system original DoF to a known state for which a universal protection mechanism (e.g., based on QZE) can be designed.
Then, the quantum superposition is swapped back to the initial DoF, restoring the original, unknown quantum state.
The protected transmission and state restoration in the initial DoF is essential since that DoF is supposed to be the operational one for the kind of protocols to be implemented, which cannot in general exploit the ancillary (decoherence-free) DoF.
Crucially, this scheme requires no prior knowledge of the input state, and allows using protection schemes agnostic to the specific details of the decoherence process, such as QZE-based methods.
This makes QSUP a universal and state-independent protection protocol that, therefore, paves the way to realistic methods for preventing decoherence in general cases.

In our specific experimental realization, we generate single-photon polarization qubits and separate their components into two distinct spatial modes of a Mach-Zehnder-like interferometer, effectively transferring the original quantum superposition of the qubit to these modes.
In each arm, the (now known) polarization undergoes engineered decoherence through controlled coupling with the photon transverse spatial distribution, simulating interaction with the environment DoFs.
Each polarization component is then protected via QZE, and subsequently recombined at the output of the interferometer, reconstructing the initial polarization-encoded qubit.

Our results show that the QSUP protocol successfully preserves both the purity and fidelity of arbitrary, unknown quantum states, regardless of the specific decoherence mechanism at play.
This demonstrates a robust, flexible, and versatile strategy for preserving quantum coherence.

\subsection*{Theoretical framework}

For the sake of simplicity, let us restrict ourselves to the case of a two-level state (i.e., a qubit) undergoing decoherence because of its interaction with the environment, e.g. when passing through a phase-damping quantum channel.
To model the decoherence, we consider a generic qubit in the state $|\psi\rangle = \alpha |0\rangle + \beta |1\rangle$, interacting with the environment in the initial state $\rho^E_\text{in}= \ketbra{E}{E}$.
The joint system is thus described by the density matrix $\rho_{in}=|\psi\rangle\langle \psi|\otimes \rho^E_\text{in}$, and its evolution is governed by the unitary operator
\begin{equation}
    \hat{U}(t)=\exp(-i\, \hat{H}\,t)\;,\label{unit}
\end{equation}
\noindent where we set $\hbar$=1.
This evolution entangles the qubit with the environment through the Hamiltonian
\begin{equation}
    \hat{H}=|\phi\rangle\langle\phi|\otimes \hat{H}_{\phi} + |\phi^\perp\rangle\langle\phi^\perp|\otimes \hat{H}_{\phi^\perp}\;,
    \label{ham}
\end{equation}
where $\hat{H}_{\phi} $ and $\hat{H}_{\phi^\perp}$ are distinct Hamiltonians coupling the states $|\phi\rangle$ and $|\phi^\perp\rangle$ to the specific environment states $|E_{\phi}(t)\rangle$ and $|E_{\phi^\perp}(t)\rangle$, being $\{|\phi\rangle,|\phi^\perp\rangle\}$ a basis for the qubit Hilbert space.
In this basis, the qubit state reads $|\psi\rangle=\delta |\phi\rangle +\eta |\phi^\perp\rangle$.
After a time $t$, the total state becomes $\rho (t)=|\Psi(t)\rangle\langle\Psi(t)|$, with $|\Psi(t)\rangle = \delta |\phi\rangle \otimes |E_{\phi}(t)\rangle + \eta |\phi^\perp\rangle \otimes |E_{\phi^\perp}(t)\rangle$.
The reduced density matrix of the qubit is obtained by tracing out the environment DoFs:
\begin{equation}
\rho(t) = \text{Tr}_E[|\Psi(t)\rangle\langle \Psi(t)|] =
\begin{pmatrix}
|\delta|^2 & \delta \eta^* \langle E_{\phi^\perp}(t)|E_{\phi}(t)\rangle \\
\delta^* \eta \langle E_{\phi}(t)|E_{\phi^\perp}(t)\rangle & |\eta|^2
\end{pmatrix}\;, \label{decstate}
\end{equation}
\noindent with $\langle E_\phi(t)|E_{\phi^\perp}(t)\rangle\in\mathbb{C}$.
When $\langle E_\phi(t)|E_{\phi^\perp}(t)\rangle\neq1$, the phase and modulus of the off-diagonal terms in Eq. \eqref{decstate} might change, causing decoherence.
For example, when $\langle E_\phi(t)|E_{\phi^\perp}(t)\rangle = e^{i\Delta\phi}$ a phase rotation is induced between the two components of the qubit, mapping the original input state onto a different (although still pure) state of the Bloch sphere.
Conversely, for $\langle E_\phi(t)|E_{\phi^\perp}(t)\rangle = 0$ the qubit state becomes a classical mixture, completely losing its quantum traits.

\begin{figure}[b]
    \centering
    \includegraphics[width=0.8\linewidth]{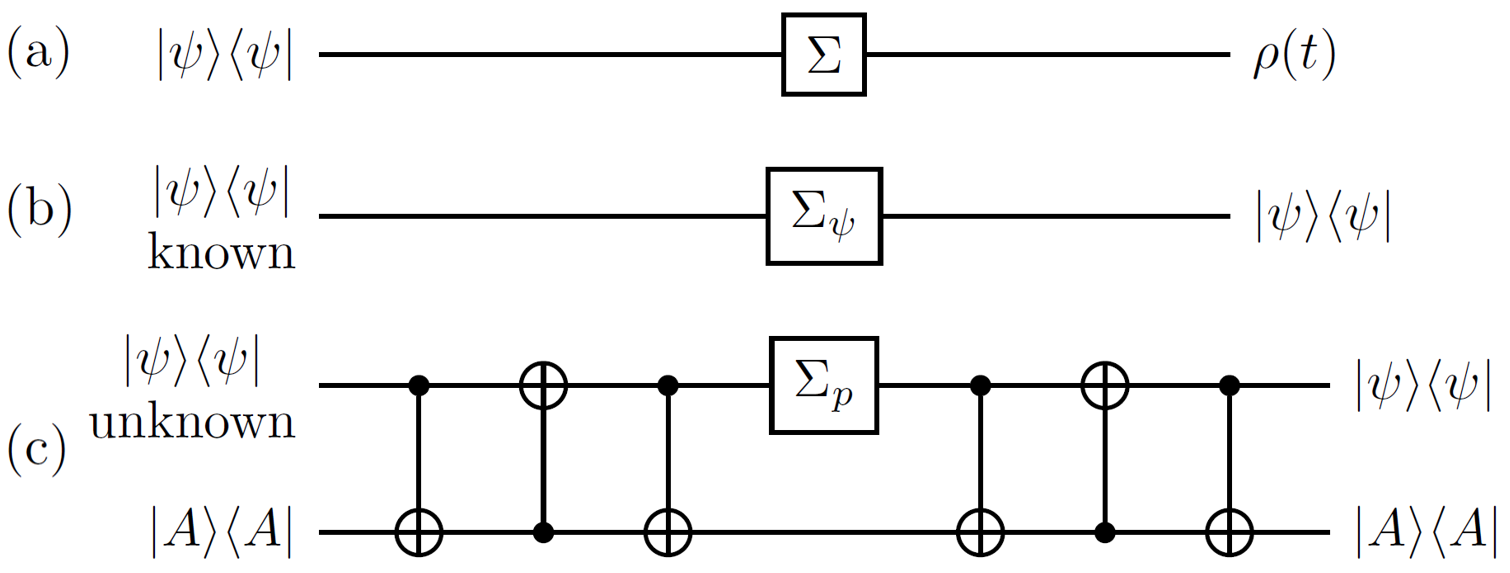}
    \caption{\textbf{Quantum State Universal Protection protocol.} Quantum circuit representation with CPTP and CP maps: (a) $\Sigma$, describing the evolution of a qubit through a decoherence channel; (b) $\Sigma_\psi$, modeling the evolution of a known qubit $\ket\psi$ through a decoherence channel with its state protected via the Quantum Zeno Effect (QZE); (c) $\Sigma_p$, realizing the protection of an unknown qubit state using the QSUP protocol with general protection map.}
    \label{scheme}
\end{figure}

The same qubit evolution can be described in terms of quantum operations, i.e. without including the environment DoF, as depicted in Fig. \ref{scheme}\textcolor{blue}{(a)}.
With this formalism, the evolution of the qubit is described by a completely-positive (CP) trace-preserving (TP) map $\Sigma$:
\begin{equation}
\rho(t) = \Sigma\left(|\psi\rangle\langle \psi|\right) = \sum_{i} \hat{K}_i(t)\, |\psi\rangle\langle \psi|\, \hat{K}_i^\dagger(t)\;, \label{krausform}
\end{equation}
\noindent where $\{\hat{K}_i(t)\}$ is the set of Kraus operators acting on the qubit and satisfy the relation $\sum_i \hat{K}_i^\dagger(t) \hat{K}_i(t) = \hat{\mathds{1}}$.

To counteract this effect, one can exploit, e.g., QZE.
Consider the same system-environment overall evolution, but now the qubit is observed (measured) $n$ times during the evolution time $t$ via projective measurements along its initial state $|\psi\rangle$.
This measurement can be described by the operator $\hat{\Pi}_{\psi} \otimes \mathds{1}_E$, where $\hat{\Pi}_{\psi}=|\psi\rangle\langle\psi|$ and $\mathds{1}_E$ is the identity on the environment subsystem.
This introduces a non-unitary evolution of the system, and the density matrix at time $t$ reads:
\begin{equation}
    \rho_{\text{QZE}} (t,n) = \frac{\text{Tr}_E\left[\left((\hat{\Pi}_{\psi}\otimes \mathds{1}_E) \hat{U}(t/n)\right)^n \rho_{in} \left(\hat{U}^\dagger(t/n)(\hat{\Pi}_{\psi}\otimes \mathds{1}_E)\right)^n\right]}{p_{sur}(t,n,\psi,\phi)}\;,
\end{equation}
\noindent where the overall trace
\begin{equation}
p_{\text{sur}}(t,n,\psi,\phi)=\text{Tr}\left[\left((\hat{\Pi}_{\psi} \otimes \mathds{1}_E)\hat{U}(t/n)\right)^n \rho_{in} \left(\hat{U}^\dagger(t/n)(\hat{\Pi}_{\psi}\otimes \mathds{1}_E)\right)^n\right]\;
\end{equation}
\noindent provides the survival probability of the qubit state to the process, i.e., the probability of finding the system in the state $|\psi\rangle$ after the $n$ projections occurring during the time interval $t$.
Due to the system-environment entanglement generated by $\hat{U}(t/n)$, the projection $\hat{\Pi}_{\psi}\otimes \mathds{1}_E$ alters the state of the environment while keeping the qubit state unchanged.
After the $(k-1)$th QZE-projection stage, the overall state can be written as the (unnormalized) product state of the qubit initial state $\ket\psi$ times a modified environment state $\rho^{(k-1)}_E$.
Then, the $k$th unitary evolution $\hat U(t/n)$ entangles the qubit with the $k$-evolved environment states $\{|E_{\phi}^{(k)}(t/n)\rangle,|E_{\phi^\perp}^{(k)}(t/n)\rangle\}$.
The qubit survival probability depends on the initial state of the qubit $|\psi\rangle$, the particular interaction (characterized by $\phi$), the total interaction time $t$, and the number of measurement steps $n$ performed:
\begin{equation}
    p_{sur}(t,n,\psi,\phi)=\prod_{k=1}^n \left[ 1-2|\delta|^2|\eta|^2\left(1-\Re\left\{\left\langle E_{\phi}^{(k)}\left(t/n\right)\bigg| E_{\phi^\perp}^{(k)}\left(t/n\right)\right\rangle\right\}\right)\right],
    \label{eq:psur1}
\end{equation}
\noindent being $\Re\{\cdots\}$ the real part and $|\delta|^2=1-|\eta|^2=|\braket{\phi}{\psi}|^2$.\\
In the limit of frequent observations, one finds that $\lim\limits_{n \to \infty}\Re\left\{\left\langle E_{\phi}^{(k)}(t/n)\bigg| E_{\phi^\perp}^{(k)}(t/n)\right\rangle\right\}=1$, so $\lim\limits_{n \to \infty}p_{\text{sur}}(t,n,\psi,\phi)=1$.
This means that the QZE projections keep the system state substantially unaltered, effectively decoupling it from the environment.
When a finite number of projections is performed, the survival probability decreases, but the output qubit state still maintains maximal fidelity and purity, i.e., $\mathcal{F}=\text{Tr}[\hat{\Pi}_{\psi}\, \rho_{\text{QZE}} (t,n)]=1$ and $\mathcal{P}=\text{Tr}[(\rho_{\text{QZE}}(t,n))^2]=1$.

This type of evolution is shown in terms of a CP map $\Sigma_\psi$ in Fig. \ref{scheme}\textcolor{blue}{(b)}.
Since now the projections cause some losses, the map is not trace preserving.
Therefore, the system evolution can be described just by the Kraus operator:
\begin{equation}
    \hat{K}_\psi = \sqrt{p_{\text{sur}}(t,n,\psi,\phi)} |\psi\rangle\langle\psi|.
    \label{QZEmap}
\end{equation}
\noindent In that case, it is straightforward to show that, when the survival probability is close to unity, the quantum channel map becomes close to the identity, as $\hat{K}_\psi$ approaches the projector $\hat{\Pi}_\psi$.

Although QZE can be used to design a protection protocol, it relies on a strong assumption: the initial state of the qubit must be known in advance in order to determine the correct projection.
This implies that it cannot be applied to a generic, unknown quantum state, or to a case where different states might travel through the channel, strongly limiting its practical interest.
Our QSUP protocol furnishes a solution to this problem, performing a universal protection by embedding the system in a larger Hilbert space, as schematically illustrated in Fig. \ref{scheme}\textcolor{blue}{(c)}.
By exploiting an ancillary, decoherence-free DoF prepared by the experimentalist, the unknown qubit state can be swapped to this DoF, bringing the original qubit state in a known state that can then be protected via standard QZE projections (or using any other protection protocol).
After the protection, the original state can be recovered by performing a reverse-swap from the ancillary state.
In our QSUP protocol, one includes for the qubit to be protected a known ancillary state $|A\rangle\langle A|$, being $\{|A\rangle,|B\rangle\}$ a basis for its Hilbert space, in the initial state, as shown in Fig. \ref{scheme}\textcolor{blue}{(c)}:
\begin{equation}\label{eq:rho_in}
    \rho_{\text{in}}= |\psi\rangle\langle \psi|\otimes |A\rangle\langle A|
\end{equation}
\noindent Then, the swap operation allows transferring the qubit superposition to the ancillary space (e.g., a further DoF less affected by decoherence, or an ancillary system evolving in a decoherence-free subspace \cite{Zanardi1997,Lidar1998}):
\begin{equation}
\rho_{\text{SWAP}} = (\hat{\text{SWAP}})\rho_{\text{in}}(\hat{\text{SWAP}})^\dagger= |\xi\rangle\langle\xi| \otimes |\psi_a\rangle\langle \psi_a|,
\label{swap}
\end{equation}
\noindent where $|\psi_a\rangle=\alpha |A\rangle + \beta |B\rangle$.
Now, the pure state $|\xi\rangle$, known to the experimentalist, can be preserved using a
general protection scheme, while passing through a decoherence channel, evolving according to the CP map $\Sigma_{\psi=\xi}$.
When one uses QZE as the protection mechanism, the map is described by the Kraus operator in Eq. \eqref{QZEmap}, and the survival probability 
will no longer depend on the initial, unknown state.
The (unnormalized) quantum state of the system now becomes:
\begin{equation}
    \rho_0 (t) = \hat{K}_\xi |\xi\rangle \langle\xi| \hat{K}_\xi^\dagger  \otimes |\psi_a\rangle\langle \psi_a|=p_{sur}^{\text{QSUP}}(t,n,\xi,\phi) |\xi\rangle \langle\xi| \otimes |\psi_a\rangle\langle\psi_a|\;,
\end{equation}
\noindent with:
\begin{equation}
    p_{sur}^{\text{QSUP}}(t,n,\xi,\phi)=\prod_{k=1}^n \left[ 1-2|\delta'|^2|\eta'|^2\left(1-\Re\left\{\left\langle E_{\phi}^{(k)}\left(t/n\right)\bigg| E_{\phi^\perp}^{(k)}\left(t/n\right)\right\rangle\right\}\right)\right]
    \label{eq:psur2}
\end{equation}
\noindent and $|\delta'|^2=1-|\eta'|^2=|\braket{\phi}{\xi}|^2$.
Finally, another swap operation allows restoring the qubit superposition in its original DoF, effectively protecting the quantum state of the system with probability $p_{sur}^{\text{QSUP}}(t,n,\xi,\phi)$, achieving the (unnormalized) state:
\begin{equation}
    \rho_\text{rec} (t) = (\hat{\text{SWAP}})\rho_0 (t)(\hat{\text{SWAP}})^\dagger= p_{sur}^{\text{QSUP}}(t,n,\xi,\phi) \rho_\text{in}\;.
    \label{swap2}
\end{equation}
\noindent As in the standard QZE protection protocol \cite{Piacentini2017,Pia2021,Virzi2024}, it can be shown that, as one goes to the constant monitoring, the survival probability approaches unity.
In the non-asymptotic regime, instead, $p_{sur}^{\text{QSUP}}(t,n,\xi,\phi)$ decreases similarly to the standard case, now depending on the $\{\ket\phi,\ket{\phi^\perp}\}$ decoherence basis and the chosen state $\ket\xi$.
Nevertheless, the final normalized state retains maximal fidelity and purity with respect to the initial unknown state $|\psi\rangle$, regardless of the particular decoherence process.
The QSUP protocol thus enables the protection of arbitrary unknown states from decoherence, requiring any prior knowledge on the quantum state, and being independent of the particular unitary evolution coupling the system to the environment.

\subsection*{Experimental implementation}

We demonstrate practical use of our QSUP protocol by implementing it in a free-space decoherence-inducing quantum channel (DIQC) transmitting single-photon qubits encoded in polarization.
To asses the universality of the protocol, we observe the Fidelity $\mathcal F$, Purity $\mathcal P$ and survival probability $p_{sur}^{\mathrm{QSUP}}$ of the QSUP-transmitted qubits for different initial states and qubit-DIQC interaction Hamiltonians of the form of Eq. \eqref{ham}.

Additionally, we compare its performance with the one obtained considering a DIQC affected by the same decoherence, but with no QSUP.
Both the setups are schematically illustrated, respectively, in the lower and upper part of Fig. \ref{fig:setup} (a detailed description of the experimental implementation is provided in the Supplementary Materials).

Both schemes of Fig. \ref{fig:setup} have three regions: Alice's lab (the sender part, preparing the qubit), the DIQC and Bob's lab (the receiver part, that measures the qubit sent by Alice).
Alice encodes the qubit in the polarization DoF of a heralded single photon, preparing it in the (linear) polarization state $\ket\psi=\cos(\psi)\ket H + \sin(\psi)\ket V$, and sending it through the DIQC.
At the DIQC exit, Bob receives the qubit and measure it with a (polarization) quantum state tomography apparatus, reconstructing the density matrix of the received quantum state.
For what concerns the DIQC, in principle our theoretical framework involves a continuous qubit-environment interaction inducing decoherence on the qubit, with QZE-projections occurring at definite time instants when QSUP is implemented.
This is equivalent to a situation in which a series of discrete, finite decoherence events occur in the DIQC, eventually interspersed by QZE-projections.
This second scenario, much easier to realize and control, is the one chosen for our experimental implementation.

For the qubit-environment coupling basis $\{\ket\phi,\ket{\phi^\perp}\}$ within the DIQC we choose $\ket\phi=\cos(\phi) \ket H +\sin(\phi)\ket V$, with the $\phi$ angle determined by half-wave plates (HWPs) at the entrance and exit of the DIQC.
Since the decoherence impact on the qubit state depends on the $\braket{\phi}{\psi}$ ($\braket{\phi}{\xi}$) overlap when QSUP is absent (present), without any loss of generality we can set the HWPs at the entrance and exit of the DIQC at $\phi=0$, realizing a qubit-environment interaction in the $\{\ket H,\ket V\}$ basis.
This way, each decoherence block is constituted by pairs of birefringent crystals arranged in order to couple the photon polarization with the photon transverse spatial DoF $x$, generating a (tiny) spatial walk-off between the two polarization components $\ket{H}$ and $\ket{V}$, while nullifying the temporal walk-off and phase mismatch between them.
This corresponds to implementing the Hamiltonian $\hat{H}_{\text{exp}} = \gamma_x\left(\hat{\Pi}_{H} \otimes \hat{P}_x + \hat{\Pi}_{V} \otimes \mathds{1}_x\right)$, where $\gamma_x$ is the coupling strength, proportional to the thickness of the crystals and $\hat{P}_x$ is the transverse momentum of the photon along the $x$ direction, canonically conjugated to the position observable $\hat{x}$ (see Supplementary Materials for further details).

To implement the QSUP protocol, a phase-stabilized Mach-Zehnder-like interferometer (MZI) is put between Alice and Bob, with the input polarizing beam-splitter (PBS) in Alice's lab and the output PBS in Bob's lab, so that the two polarization components of the qubit travel through different paths within the same DIQC.
This way, the path acts as the ancillary DoF in the scheme of Fig. \ref{scheme}(c), evolving in a decoherence-free subspace.
A pair of HWPs, one per MZI arm, is placed after Alice's PBS to realize, together with it, the SWAP operation related to Eq. \eqref{swap}, i.e. creating the state $\ket\xi\otimes\ket{\psi_a}$, with $\ket\xi=\cos(\xi) \ket H +\sin(\xi)\ket V$.
Once the qubit and ancilla states are swapped, in both arms of the MZI the QZE-protection mechanism is realized on the qubit wavefunction components by subsequent projections onto the polarization state $\ket\xi$, by a series of polarizers interspersing the decoherence blocks.
At Bob's station, a pair of HWPs (identical to the ones of Alice's lab) followed by a PBS coherently recombine the qubit wavefunction components in Bob's lab, implementing the final SWAP operation of the protocol [Eq. \eqref{swap2}] and restoring the initial state in Eq. \eqref{eq:rho_in}.
Finally, Bob performs quantum tomographic reconstruction of the polarization state of the received qubit by means of a HWP, a quarter-wave plate (QWP) and a polarizer.
For the sake of a fair comparison, we also implement a transmission within the same DIQC without QSUP, i.e. by removing the MZI responsible for the SWAP operations and, as a consequence, the support of the ancillary DoF needed to implement the QZE-based protection.

We perform the tests with the qubit initial states $\ket\psi$, with $\psi=20^\circ,45^\circ,60^\circ$, and with the QSUP-protected states $\ket{\xi}$, with $\xi=20^\circ,45^\circ,60^\circ$.
These values are chosen to avoid cases in which the interaction negligibly affects the system.

\begin{figure}
    \centering
    \includegraphics[width=0.80\linewidth, trim={1.2cm 0 1.2cm 0},clip]{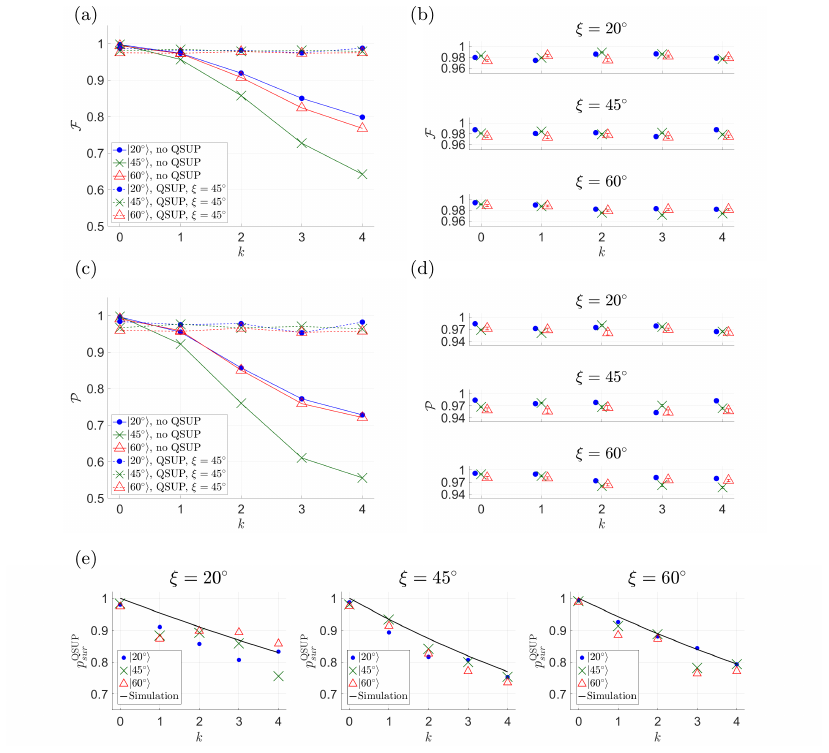}
    \caption{\textbf{Performance assessment.}
    Comparison between unprotected and protected transmission of different polarization-encoded qubit states $\ket\psi = \cos(\psi)\,\ket{H} + \sin(\psi)\,\ket{V}$ ($\psi = 20^\circ$, $45^\circ$, $60^\circ$) through a DIQC, considering different protected states $\ket\xi$ ($\xi = 20^\circ$, $45^\circ$, $60^\circ$).
    Plot (a): Fidelity $\mathcal F$ of the qubit state exiting the DIQC with the one entering it, both in the protected and unprotected cases, considering for the QSUP the worst-case scenario of maximal decoherence ($\xi=45^\circ$).
    (b) Fidelity $\mathcal F$ after the QSUP-assisted transmission for three different input states $\ket\psi$ and protected states $\ket\xi$.
    (c) Purity $\mathcal P$ of the qubit state exiting the DIQC with and without QSUP, considering for the QSUP the worst-case scenario of maximal decoherence ($\xi=45^\circ$).
    (d) Purity $\mathcal P$ after the QSUP-assisted transmission in the DIQC for three different input states and qubit-environment couplings.
    (e) Qubit survival probability $p_{sur}^{\mathrm{QSUP}}$, considering three different input states $\ket\psi$ and protected states $\ket\xi$. The theoretical simulations (black curves) include absorption-due optical losses.
}
    \label{fig:data}
\end{figure}

In Fig. \ref{fig:data}, we report the experimental results for the Fidelity $\mathcal F$, Purity $\mathcal P$, and survival probability $p_{sur}^{\mathrm{QSUP}}$ of the reconstructed qubit state after $k=0,..,4$ decoherence blocks.
Specifically, Fig. \ref{fig:data}\textcolor{blue}{(a)} and Fig. \ref{fig:data}\textcolor{blue}{(c)} compare, respectively, the qubit Fidelity and Purity obtained for the unprotected DIQC with the ones obtained for the QSUP-assisted one in the worst-case scenario, i.e. for $\xi=45^\circ$, for which the predicted $p_{sur}^\text{QSUP}$ (Eq. \ref{eq:psur2}) is minimal.
Without QSUP, $\mathcal F$ and $\mathcal P$ rapidly degrade with $k$ due to growing entanglement between the polarization (qubit DoF) and the transversal spatial distribution (environment DoF), leading to decoherence eventually causing a transition towards a mixed state.
In contrast, with QSUP both $\mathcal F$ and $\mathcal P$ remain close to unity for all $k$ values, clearly demonstrating the ability of QSUP in suppressing decoherence for all tested input states, protecting the quantum information stored in the qubit.
Moreover, we demonstrate that the protocol preserves the coherence of the qubit state encoded by Alice independently of the specific protected state $\ket\xi$, as shown in Figs. \ref{fig:data}\textcolor{blue}{(b)} and \ref{fig:data}\textcolor{blue}{(d)}.
Finally, Fig. \ref{fig:data}\textcolor{blue}{(e)} shows the decrease of the qubit survival probability $p_{sur}^{\mathrm{QSUP}}$ in Eq. \eqref{eq:psur2} for increasing $k$ for three different $\ket\xi$ states, because of the QZE-projections performed at each stage of the protocol to protect the qubit.\\
Indeed, these projections unavoidably introduce losses due to the non-negligible coupling between qubit and environment (see Supplementary Materials for details).
A larger coupling strength increases the distinguishability of the environment states, reducing the probability of projecting the polarization back into its initial state, as can be seen in Eq. \eqref{eq:psur2}.
After $k=4$ steps, Figs. \ref{fig:data}\textcolor{blue}{(a)} and \ref{fig:data}\textcolor{blue}{(c)} show that, in the worst case scenario, the unprotected state becomes close to a completely mixed one, with Fidelity and Purity approaching values related to states with no quantum coherence.
In contrast, the protected state retains high Fidelity ($\mathcal F>0.96$) and Purity ($\mathcal P>0.94$) for all the states and qubit-environment couplings considered, demonstrating that the polarization-encoded quantum information is preserved with a survival probability (including all the optical losses) still exceeding $73\%$ even in the maximal-decoherence condition.
This survival probability can be further enhanced by increasing the number of QZE-projections, thus lowering the amount of environment-induced decoherence between subsequent projections and, therefore, increasing the success probability of each projection.
It is worth stressing that, although the non-unit survival probability $p_{sur}^{\mathrm{QSUP}}$ inevitably results in a reduced amount of physical qubits exchanged between Alice and Bob, the close-to-unity Purity and Fidelity granted by our protection protocol allow delivering a much higher amount of reliable, error-free quantum information, boosting the transmission efficiency in practical decoherence-affected scenarios.

These results provide compelling experimental evidence that our protocol protects the quantum state of the system independently of the input state and the particular qubit-environment interaction, thereby confirming the universality of the QSUP approach.

\section*{Discussion}

Transmitting quantum information while preserving quantum coherence remains a fundamental challenge for the advancement of quantum technologies. Interactions with the environment in realistic quantum channels inevitably induce decoherence, rapidly degrading quantum states and the information they encode.
In this work, we have proposed and experimentally demonstrated a novel universal protocol for protecting arbitrary and unknown quantum states against decoherence.
Our strategy is universally applicable: it does not depend on the form of the system–environment coupling nor on the specific quantum state, in contrast to state-dependent protection schemes or tailored quantum error-correction codes. As such, it provides a broadly applicable solution for quantum communication, distributed quantum sensing, and quantum computation.
As a specific application, we tested our protocol on a photonic-qubit-based platform.
By comparing Fidelity, Purity, and survival probability in protected and unprotected evolutions, we show that our protocol effectively preserves quantum coherence, keeping Fidelity and Purity close to unity, whereas without protection the state would rapidly decohere and the polarization state of the qubit at the end of the channel becomes unsuitable for any quantum protocol.
In this context, the unavoidable losses introduced by our protocol actually represent only a minor cost for the near-intact preservation of the quantum information encoded in the transmitted qubits.
Beyond free-space implementations, the protocol is compatible with integrated and fiber-based photonic technologies and can be extended to other physical platforms employing ancillary degrees of freedom. Overall, our results provide a practical and general route toward state-independent decoherence suppression across quantum technologies.



\begin{figure}
    \centering
    \includegraphics[width=0.90\linewidth]{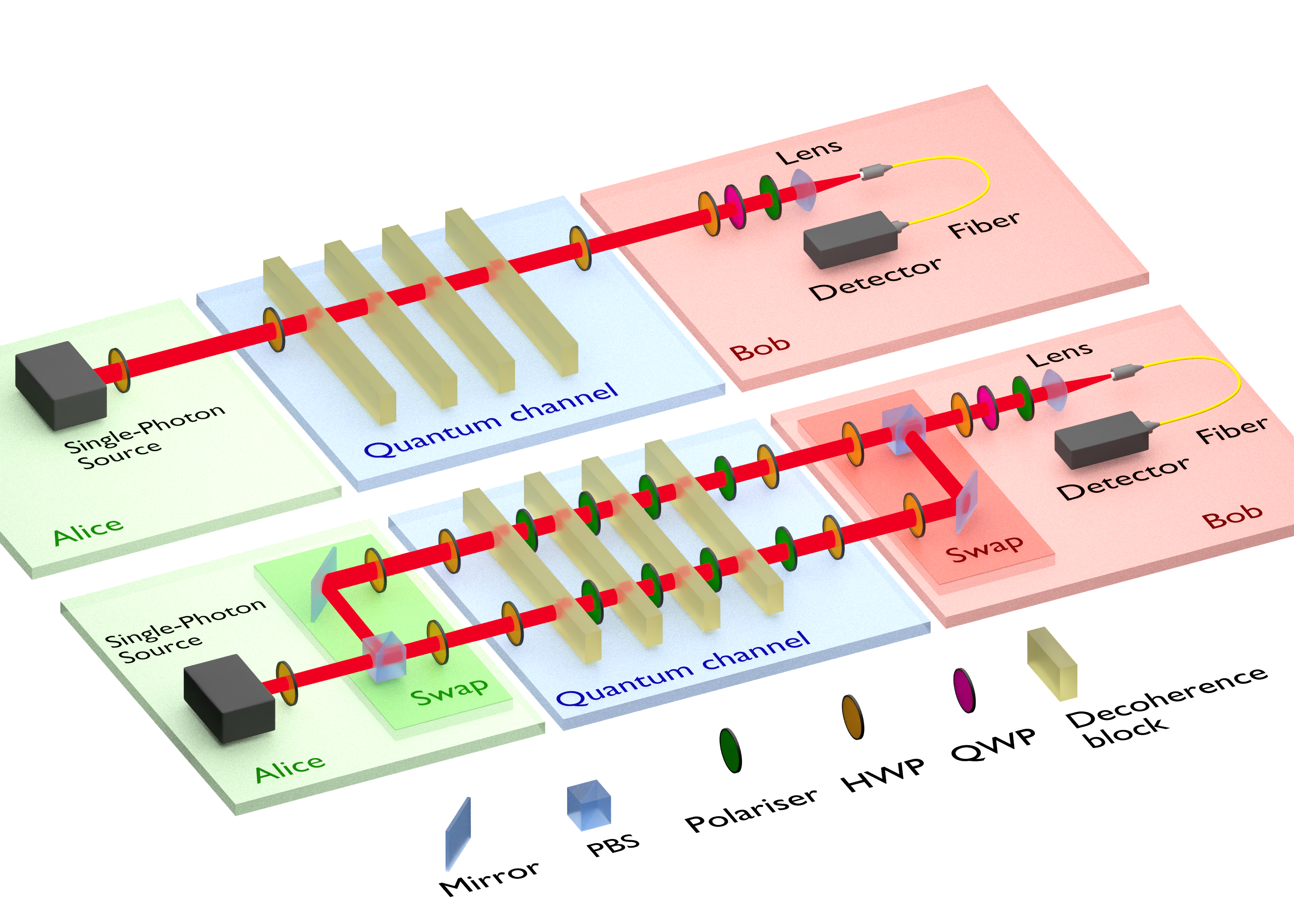}
    \caption{\textbf{Single-photon-based implementation of Quantum State Universal Protection (QSUP).}
    Experimental scheme of our quantum channel with (lower sketch) and without (upper sketch) QSUP. For the QSUP protocol, Alice encodes a single-photon qubit in its polarization DoF, sending it to a Mach-Zehnder-like interferometer (MZI).
    There, the SWAP operator between the qubit and the ancilla (here, the MZI path DoF) in Fig. \ref{scheme}(c) is realized by a polarizing beam-splitter (PBS) followed by one half-wave plate (HWP) in each arm, rotating both polarization components onto the $\ket\xi$ state to be QZE-protected.
    In the channel, two HWPs (one per arm) set the basis of the qubit-environment coupling $\{\ket\phi,\ket{\phi^\perp}\}$. Such coupling is implemented by a series of discrete birefringence-based (tiny) decoherence blocks, interspersed by polarizers performing QZE-projections onto the state $\ket\xi$ (at the channel output, two identical HWPs counter-rotate the $\{\ket\phi,\ket{\phi^\perp}\}$ basis into the computational one).
    On Bob's side, a pair of HWPs and a PBS close the MZI re-performing the SWAP, restoring the initially-encoded polarization state.
    Then, the qubit undergoes quantum state tomography by means of an apparatus constituted by a HWP, a quarter-wave plate (QWP) and a polarizer.
    The channel without QSUP [Fig. \ref{scheme}(a)] is implemented by removing the MZI as well as the polarizers for the QZE, while keeping the same decoherence-inducing elements.}
    \label{fig:setup}
\end{figure}

\clearpage 

%
\bibliography{QSUP-bibliography} 

@article{Wendin2017,
doi = {10.1088/1361-6633/aa7e1a},
url = {https://dx.doi.org/10.1088/1361-6633/aa7e1a},
year = {2017},
month = {sep},
publisher = {IOP Publishing},
volume = {80},
number = {10},
pages = {106001},
author = {Wendin, G},
title = {Quantum information processing with superconducting circuits: a review},
journal = {Reports on Progress in Physics},
}

@article{Jeremy2019,
    author = {Bruzewicz, Colin D. and Chiaverini, John and McConnell, Robert and Sage, Jeremy M.},
    title = {Trapped-ion quantum computing: Progress and challenges},
    journal = {Applied Physics Reviews},
    volume = {6},
    number = {2},
    pages = {021314},
    year = {2019},
    month = {05},
    issn = {1931-9401},
    doi = {10.1063/1.5088164},
    url = {https://doi.org/10.1063/1.5088164},
}

@article{Gerard2003,
author = {MacFarlane, A. G. J.  and Dowling, Jonathan P.  and Milburn, Gerard J. },
title = {Quantum technology: the second quantum revolution},
journal = {Philosophical Transactions of the Royal Society of London. Series A: Mathematical, Physical and Engineering Sciences},
volume = {361},
number = {1809},
pages = {1655-1674},
year = {2003},
doi = {10.1098/rsta.2003.1227}
}

@article{Kim2017,
title = {From quantum optics to quantum technologies},
journal = {Progress in Quantum Electronics},
volume = {54},
pages = {2-18},
year = {2017},
note = {Special issue in honor of the 70th birthday of Professor Sir Peter Knight FRS},
issn = {0079-6727},
doi = {https://doi.org/10.1016/j.pquantelec.2017.06.002},
url = {https://www.sciencedirect.com/science/article/pii/S0079672717300186},
author = {Dan Browne and Sougato Bose and Florian Mintert and M.S. Kim},
}

@article{Viola1998,
  title = {Dynamical suppression of decoherence in two-state quantum systems},
  author = {Viola, Lorenza and Lloyd, Seth},
  journal = {Phys. Rev. A},
  volume = {58},
  issue = {4},
  pages = {2733--2744},
  numpages = {0},
  year = {1998},
  month = {Oct},
  publisher = {American Physical Society},
  doi = {10.1103/PhysRevA.58.2733},
  url = {https://link.aps.org/doi/10.1103/PhysRevA.58.2733}
}

@article{Viola1999,
  title = {Dynamical Decoupling of Open Quantum Systems},
  author = {Viola, Lorenza and Knill, Emanuel and Lloyd, Seth},
  journal = {Phys. Rev. Lett.},
  volume = {82},
  issue = {12},
  pages = {2417--2421},
  numpages = {0},
  year = {1999},
  month = {Mar},
  publisher = {American Physical Society},
  doi = {10.1103/PhysRevLett.82.2417},
  url = {https://link.aps.org/doi/10.1103/PhysRevLett.82.2417}
}

@article{Kofman2001,
  title = {Universal Dynamical Control of Quantum Mechanical Decay: Modulation of the Coupling to the Continuum},
  author = {Kofman, A. G. and Kurizki, G.},
  journal = {Phys. Rev. Lett.},
  volume = {87},
  issue = {27},
  pages = {270405},
  numpages = {4},
  year = {2001},
  month = {Dec},
  publisher = {American Physical Society},
  doi = {10.1103/PhysRevLett.87.270405},
  url = {https://link.aps.org/doi/10.1103/PhysRevLett.87.270405}
}

@article{Kofman2000,
  author    = {Kofman, A. G. and Kurizki, G.},
  title     = {Acceleration of quantum decay processes by frequent observations},
  journal   = {Nature},
  volume    = {405},
  number    = {6786},
  pages     = {546--550},
  year      = {2000},
  month     = jun,
  doi       = {10.1038/35014537},
  url       = {https://doi.org/10.1038/35014537},
  issn      = {1476-4687}
}

@article{Souza2012,
  title = {Experimental protection of quantum gates against decoherence and control errors},
  author = {Souza, Alexandre M. and \'Alvarez, Gonzalo A. and Suter, Dieter},
  journal = {Phys. Rev. A},
  volume = {86},
  issue = {5},
  pages = {050301},
  numpages = {4},
  year = {2012},
  month = {Nov},
  publisher = {American Physical Society},
  doi = {10.1103/PhysRevA.86.050301},
  url = {https://link.aps.org/doi/10.1103/PhysRevA.86.050301}
}

@article{Virzi2022,
  title = {Quantum Zeno and Anti-Zeno Probes of Noise Correlations in Photon Polarization},
  author = {Virz\`{\i}, Salvatore and Avella, Alessio and Piacentini, Fabrizio and Gramegna, Marco and Opatrn\'y, Tom\'a\ifmmode \check{s}\else \v{s}\fi{} and Kofman, Abraham G. and Kurizki, Gershon and Gherardini, Stefano and Caruso, Filippo and Degiovanni, Ivo Pietro and Genovese, Marco},
  journal = {Phys. Rev. Lett.},
  volume = {129},
  issue = {3},
  pages = {030401},
  numpages = {7},
  year = {2022},
  month = {Jul},
  publisher = {American Physical Society},
  doi = {10.1103/PhysRevLett.129.030401},
  url = {https://link.aps.org/doi/10.1103/PhysRevLett.129.030401}
}

@article{kim2012,
  title = {Protecting entanglement from decoherence using weak measurement and quantum measurement reversal},
  author = {Yong-Su Kim, Jong-Chan Lee, Osung Kwon and Yoon-Ho Kim},
  journal = {Nature Physics},
  volume = {8},
  pages = {117–120},
  year = {2012},
  doi = {https://doi.org/10.1038/nphys2178},
}

@article{OBrien2009,
  author    = {Jeremy L. O'Brien and Akira Furusawa and Jelena Vučković},
  title     = {Photonic quantum technologies},
  journal   = {Nature Photonics},
  volume    = {3},
  number    = {12},
  pages     = {687--695},
  year      = {2009},
  doi       = {10.1038/nphoton.2009.229},
  url       = {https://doi.org/10.1038/nphoton.2009.229}
}

@article{Imre2019,
title = {A Survey on quantum computing technology},
journal = {Computer Science Review},
volume = {31},
pages = {51-71},
year = {2019},
issn = {1574-0137},
doi = {https://doi.org/10.1016/j.cosrev.2018.11.002},
url = {https://www.sciencedirect.com/science/article/pii/S1574013718301709},
author = {Laszlo Gyongyosi and Sandor Imre},
}

@article{Kumar2024,
  author={Zhang, Peiying and Chen, Ning and Shen, Shigen and Yu, Shui and Wu, Sheng and Kumar, Neeraj},
  journal={IEEE Wireless Communications}, 
  title={Future Quantum Communications and Networking: A Review and Vision}, 
  year={2024},
  volume={31},
  number={1},
  pages={141-148},
  doi={10.1109/MWC.012.2200295}}

@article{Cappellaro2017,
  title = {Quantum sensing},
  author = {Degen, C. L. and Reinhard, F. and Cappellaro, P.},
  journal = {Rev. Mod. Phys.},
  volume = {89},
  issue = {3},
  pages = {035002},
  numpages = {39},
  year = {2017},
  month = {Jul},
  publisher = {American Physical Society},
  doi = {10.1103/RevModPhys.89.035002},
  url = {https://link.aps.org/doi/10.1103/RevModPhys.89.035002}
}

@article{Zurek2003,
  title = {Decoherence, einselection, and the quantum origins of the classical},
  author = {Zurek, Wojciech Hubert},
  journal = {Rev. Mod. Phys.},
  volume = {75},
  issue = {3},
  pages = {715--775},
  numpages = {0},
  year = {2003},
  month = {May},
  publisher = {American Physical Society},
  doi = {10.1103/RevModPhys.75.715},
  url = {https://link.aps.org/doi/10.1103/RevModPhys.75.715}
}

@article{Sudarshan1977,
    author = {Misra, B. and Sudarshan, E. C. G.},
    title = {The Zeno’s paradox in quantum theory},
    journal = {Journal of Mathematical Physics},
    volume = {18},
    number = {4},
    pages = {756-763},
    year = {1977},
    month = {04},
    issn = {0022-2488},
    doi = {10.1063/1.523304},
    url = {https://doi.org/10.1063/1.523304}
}

@article{Wineland1990,
  title = {Quantum Zeno effect},
  author = {Itano, Wayne M. and Heinzen, D. J. and Bollinger, J. J. and Wineland, D. J.},
  journal = {Phys. Rev. A},
  volume = {41},
  issue = {5},
  pages = {2295--2300},
  numpages = {0},
  year = {1990},
  month = {Mar},
  publisher = {American Physical Society},
  doi = {10.1103/PhysRevA.41.2295},
  url = {https://link.aps.org/doi/10.1103/PhysRevA.41.2295}
}

@article{Tasaki2006,
author = {Facchi, Paolo and Nakazato, Hiromichi and Pascazio, Saverio and Tasaki, Shuichi},
title = {CONTROL OF DECOHERENCE VIA QUANTUM ZENO SUBSPACES},
journal = {International Journal of Modern Physics B},
volume = {20},
number = {11n13},
pages = {1408-1420},
year = {2006},
doi = {10.1142/S0217979206034017},
URL = {https://doi.org/10.1142/S0217979206034017
}}

@article{Kondo2016,
doi = {10.1088/1367-2630/18/1/013033},
url = {https://dx.doi.org/10.1088/1367-2630/18/1/013033},
year = {2016},
month = {jan},
publisher = {IOP Publishing},
volume = {18},
number = {1},
pages = {013033},
author = {Kondo, Yasushi and Matsuzaki, Yuichiro and Matsushima, Kei and Filgueiras, Jefferson G},
title = {Using the quantum Zeno effect for suppression of decoherence},
journal = {New Journal of Physics},
}

@article{Facchi2008,
doi = {10.1088/1751-8113/41/49/493001},
url = {https://dx.doi.org/10.1088/1751-8113/41/49/493001},
year = {2008},
month = {oct},
publisher = {},
volume = {41},
number = {49},
pages = {493001},
author = {Facchi, P and Pascazio, S},
title = {Quantum Zeno dynamics: mathematical and physical aspects},
journal = {Journal of Physics A: Mathematical and Theoretical},
}

@article{Genovese2016,
doi = {10.1088/2040-8978/18/7/073002},
url = {https://doi.org/10.1088/2040-8978/18/7/073002},
year = {2016},
month = {jun},
publisher = {IOP Publishing},
volume = {18},
number = {7},
pages = {073002},
author = {Genovese, Marco},
title = {Real applications of quantum imaging},
journal = {Journal of Optics}
}

@article{Petrini2020,
author = {Petrini, Giulia and Moreva, Ekaterina and Bernardi, Ettore and Traina, Paolo and Tomagra, Giulia and Carabelli, Valentina and Degiovanni, Ivo Pietro and Genovese, Marco},
title = {Is a Quantum Biosensing Revolution Approaching? Perspectives in NV-Assisted Current and Thermal Biosensing in Living Cells},
journal = {Advanced Quantum Technologies},
volume = {3},
number = {12},
pages = {2000066},
doi = {https://doi.org/10.1002/qute.202000066},
url = {https://advanced.onlinelibrary.wiley.com/doi/abs/10.1002/qute.202000066},
eprint = {https://advanced.onlinelibrary.wiley.com/doi/pdf/10.1002/qute.202000066},
year = {2020}
}

@article {Genovese2010,
author = "Genovese, Marco",
title = "Interpretations of Quantum Mechanics and Measurement Problem",
journal = "Advanced Science Letters",
volume = "3",
number = "3",
year = "2010",
itemtype = "article",
issn = "1936-6612",
date ="2010-09-01",
pages = "249-258",
url = "https://www.ingentaconnect.com/content/asp/asl/2010/00000003/00000003/art00001"
}

@article{Zanardi1997,
  title = {Noiseless Quantum Codes},
  author = {Zanardi, P. and Rasetti, M.},
  journal = {Phys. Rev. Lett.},
  volume = {79},
  issue = {17},
  pages = {3306--3309},
  numpages = {0},
  year = {1997},
  month = {Oct},
  publisher = {American Physical Society},
  doi = {10.1103/PhysRevLett.79.3306},
  url = {https://link.aps.org/doi/10.1103/PhysRevLett.79.3306}
}

@article{Lidar1998,
  title = {Decoherence-Free Subspaces for Quantum Computation},
  author = {Lidar, D. A. and Chuang, I. L. and Whaley, K. B.},
  journal = {Phys. Rev. Lett.},
  volume = {81},
  issue = {12},
  pages = {2594--2597},
  numpages = {0},
  year = {1998},
  month = {Sep},
  publisher = {American Physical Society},
  doi = {10.1103/PhysRevLett.81.2594},
  url = {https://link.aps.org/doi/10.1103/PhysRevLett.81.2594}
}

@article{Vitali1999,
  title = {Using parity kicks for decoherence control},
  author = {Vitali, D. and Tombesi, P.},
  journal = {Phys. Rev. A},
  volume = {59},
  issue = {6},
  pages = {4178--4186},
  numpages = {0},
  year = {1999},
  month = {Jun},
  publisher = {American Physical Society},
  doi = {10.1103/PhysRevA.59.4178},
  url = {https://link.aps.org/doi/10.1103/PhysRevA.59.4178}
}

@Article{Piacentini2017,
author={Piacentini, Fabrizio and Avella, Alessio and Rebufello, Enrico and Lussana, Rudi and Villa, Federica and Tosi, Alberto and Gramegna, Marco and Brida, Giorgio and Cohen, Eliahu and Vaidman, Lev and Degiovanni, Ivo Pietro and Genovese, Marco},
title={Determining the quantum expectation value by measuring a single photon},
journal={Nat. Phys.},
year={2017},
volume={13},
number={12},
pages={1191-1194},
issn={1745-2481},
doi={10.1038/nphys4223},
url={https://doi.org/10.1038/nphys4223}
}

@article{Virzi2024,
  title = {Sensing microscopic noise events by frequent quantum measurements},
  author = {Virz\`{\i}, Salvatore and Knoll, Laura T. and Avella, Alessio and Piacentini, Fabrizio and Gherardini, Stefano and Gramegna, Marco and Kurizki, Gershon and Kofman, Abraham G. and Degiovanni, Ivo Pietro and Genovese, Marco and Caruso, Filippo},
  journal = {Phys. Rev. Appl.},
  volume = {21},
  issue = {3},
  pages = {034014},
  numpages = {9},
  year = {2024},
  month = {Mar},
  publisher = {American Physical Society},
  doi = {10.1103/PhysRevApplied.21.034014},
  url = {https://link.aps.org/doi/10.1103/PhysRevApplied.21.034014}
}

@book{ParisRehacek2004,
  editor    = {Matteo G. A. Paris and Jaroslav Řeháček},
  title     = {Quantum State Estimation},
  series    = {Lecture Notes in Physics},
  volume    = {649},
  publisher = {Springer Berlin Heidelberg},
  address   = {Berlin, Heidelberg},
  year      = {2004},
  isbn      = {978-3-540-22329-0},
  pages     = {xiii + 520},
  doi       = {10.1007/b98673}
}

@article{Giov2006,
  title = {Quantum Metrology},
  author = {Giovannetti, Vittorio and Lloyd, Seth and Maccone, Lorenzo},
  journal = {Phys. Rev. Lett.},
  volume = {96},
  issue = {1},
  pages = {010401},
  numpages = {4},
  year = {2006},
  month = {Jan},
  publisher = {American Physical Society},
  doi = {10.1103/PhysRevLett.96.010401},
  url = {https://link.aps.org/doi/10.1103/PhysRevLett.96.010401}
}

@article{Gen2021,
    author = {Genovese, M.},
    title = {Experimental quantum enhanced optical interferometry},
    journal = {AVS Quantum Science},
    volume = {3},
    number = {4},
    pages = {044702},
    year = {2021},
    month = {11},
    issn = {2639-0213},
    doi = {10.1116/5.0062114},
    url = {https://doi.org/10.1116/5.0062114},
}

@Article{Pia2021,
AUTHOR = {Rebufello, Enrico and Piacentini, Fabrizio and Avella, Alessio and Lussana, Rudi and Villa, Federica and Tosi, Alberto and Gramegna, Marco and Brida, Giorgio and Cohen, Eliahu and Vaidman, Lev and Degiovanni, Ivo Pietro and Genovese, Marco},
TITLE = {Protective Measurement—A New Quantum Measurement Paradigm: Detailed Description of the First Realization},
JOURNAL = {Applied Sciences},
VOLUME = {11},
YEAR = {2021},
NUMBER = {9},
ARTICLE-NUMBER = {4260},
URL = {https://www.mdpi.com/2076-3417/11/9/4260},
ISSN = {2076-3417},
DOI = {10.3390/app11094260}
}

@article{reb2021,
doi = {10.1038/s41377-021-00539-0},
url = {https://doi.org/10.1038/s41377-021-00539-0},
year = {2021},
month = {may},
publisher = {Springer Nature},
volume = {10},
number = {1},
pages = {106},
author = {Rebufello, Enrico and Piacentini, Fabrizio and Avella, Alessio and Souza, Muriel A. de and Gramegna, Marco and Dziewior, Jan and Cohen, Eliahu and Vaidman, Lev and Degiovanni, Ivo Pietro and Genovese, Marco},
title = {Anomalous weak values via a single photon detection},
journal = {Light: Science \& Applications}
}

@article{reb2025,
author = {Rebufello, Enrico and Piacentini, Fabrizio and Avella, Alessio and de Souza, Muriel A. and Gramegna, Marco and Lussana, Rudi and Villa, Federica and Dziewior, Jan and Cohen, Eliahu and Vaidman, Lev and Degiovanni, Ivo Pietro and Genovese, Marco},
title = {Robust Weak Measurements with Certified Single Photons},
journal = {Advanced Quantum Technologies},
volume = {8},
number = {7},
pages = {2400482},
doi = {https://doi.org/10.1002/qute.202400482},
url = {https://advanced.onlinelibrary.wiley.com/doi/abs/10.1002/qute.202400482},
eprint = {https://advanced.onlinelibrary.wiley.com/doi/pdf/10.1002/qute.202400482},
year = {2025}
}
\bibliographystyle{ieeetr}

%
%
%
%
%
%


\section*{Acknowledgments}



The results presented in this article had been achieved also in the context of the following projects: QUID (QUantum Italy Deployment) and EQUO (European QUantum ecOsystems), which are funded by the European Commission in the Digital Europe Programme under the grant agreements number 101091408 and 101091561; QU-TEST, which had received funding from the European Union's Horizon Europe under the grant agreement number 101113901. This work was also funded by the project 23NRM04 NoQTeS, which received funding from the European Partnership on Metrology, co-financed from the European Union's Horizon Europe Research and Innovation Programme and by the Participating States.


F.A., S.V., A.A., F.P., I.P.D., M.Gr. and M.Ge. planned the experiment, based on the original idea proposed by A.A. The experimental realization was carried out by F.A., S.V., F.D. and D.A., under the supervision of A.A., F.P. and I.P.D., in the laboratories led by M.Ge.
All authors contributed to the discussion of the results, and to the drafting of the manuscript.

\newpage


\renewcommand{\thefigure}{S\arabic{figure}}
\renewcommand{\thetable}{S\arabic{table}}
\renewcommand{\theequation}{S\arabic{equation}}
\renewcommand{\thepage}{S\arabic{page}}
\setcounter{figure}{0}
\setcounter{table}{0}
\setcounter{equation}{0}
\setcounter{page}{1} 


\begin{center}
\section*{Supplementary Materials}

	Francesco~Atzori$^{1,2}$,
	Salvatore~Virzì$^{1}$,
    Francesco~Devecchi$^{3}$,
    Domenico~Abbondandolo$^{1,2}$,
    Alessio~Avella$^{1\ast}$,
    Fabrizio~Piacentini$^{1}$,
    Marco~Gramegna$^{1}$,
    Ivo~Pietro~Degiovanni$^{1}$,
	Marco~Genovese$^{1,4}$

	\small$^{1}$ Istituto Nazionale di Ricerca Metrologica, Strada delle Cacce, 91, Torino, 10135, Italy.\\
	\small$^{2}$Politecnico di Torino, Corso Castelfidardo, 39, Torino, 10129, Italy.\\
    \small$^{3}$Università degli Studi di Torino, via P. Giuria 1, Torino, 10125, Italy.\\
    \small$^{4}$Istituto Nazionale di Fisica Nucleare, via P. Giuria 1, Torino, 10125, Italy.\\
	\small$^\ast$Corresponding author. Email: a.avella@inrim.it

\end{center}




\newpage



\subsection*{Experimental setup: technical details}

\begin{figure} 
	\centering
	\includegraphics[width=1 \textwidth]{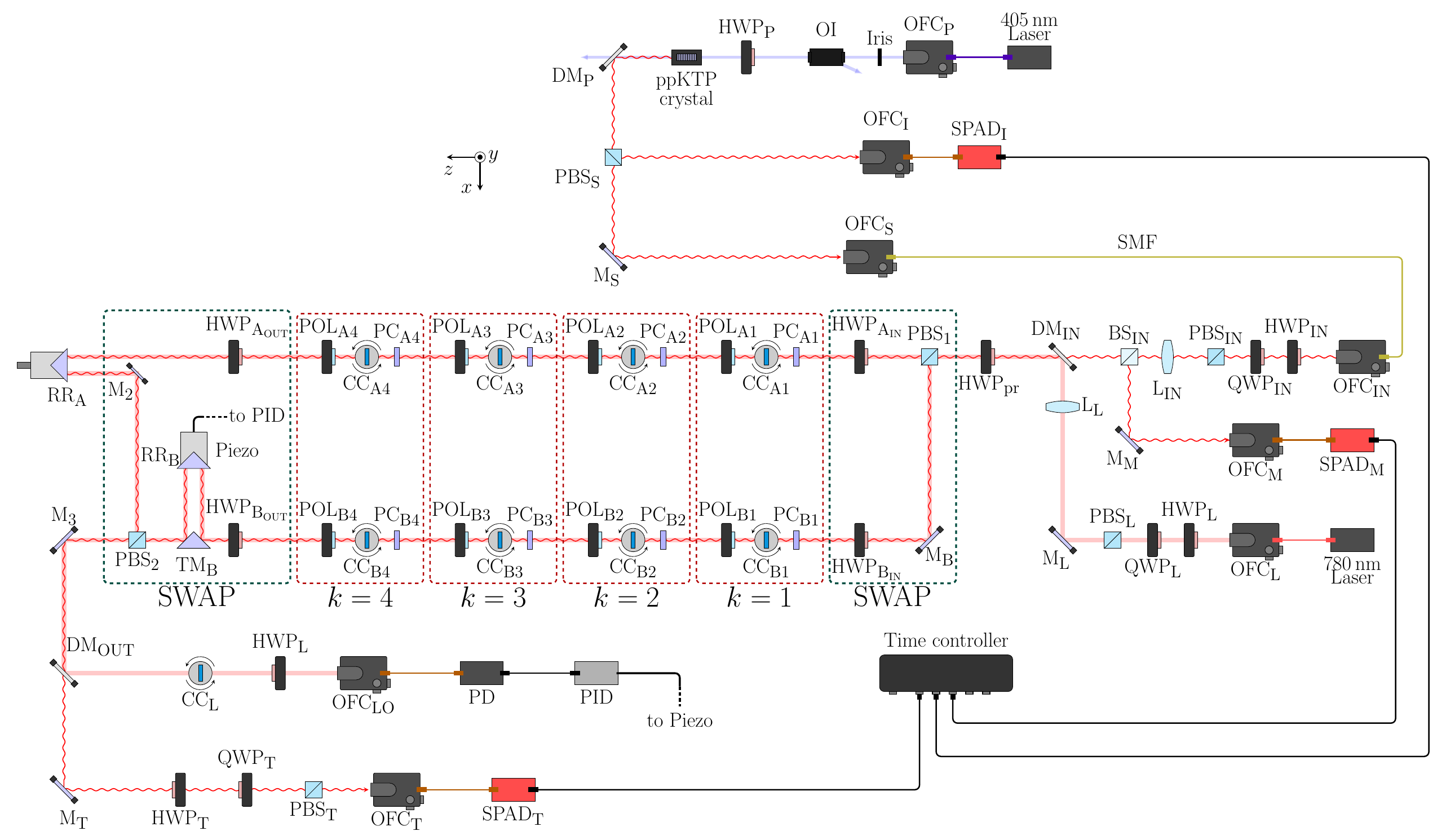} 

	\caption{\textbf{Detailed experimental scheme.}
		Heralded single photons at 810 nm are generated via degenerate type-II SPDC occurring within a ppKTP crystal.
    The L$_\text{IN}$ lens collimates the photons along the DIQC, with a 50:50 BS (BS$_\text{IN}$) directing half of the photons to a monitor channel to track source stability during acquisition.
    For the QSUP protocol implementation, the combination of PBS$_1$ and HWP$_{\text{A}_\text{IN},\text{B}_\text{IN}}$ performs a SWAP from polarization modes to ancillary path modes of the MZI.
    Up to $N=4$ decoherence blocks per arm, implemented with birefringent crystal pairs (PC$_i$, CC$_i$), couple the photon polarization to the transverse spatial distribution (which simulates the environment DoF), while the POL$_i$ polarizers realize the QZE-protection.
    Finally, HWP$_{\text{A}_\text{OUT},\text{B}_\text{OUT}}$ and PBS$_2$ recombine the spatial modes restoring the initial qubit polarization state, which is reconstructed via a quantum state tomography apparatus (HWP$_\text{T}$, QWP$_\text{T}$, PBS$_\text{T}$).
    A stabilization laser at 780 nm, overlapped with the heralded-photon path and then split from it via dichroic mirrors (DM$_\text{IN}$ and DM$_\text{OUT}$, respectively), controls and stabilizes the interferometer.
    OFC: optical fiber coupler; OI: optical isolator; HWP: half-wave plate; QWP: quarter-wave plate; M: mirror; ppKTP: periodically-poled potassium titanyl phosphate; PBS: polarizing beam splitter; DM: dichroic mirror; SMF: single-mode fiber; L: lens; SPAD: single-photon avalanche diode; POL: polarizer; PID: proportional-integral-derivative controller; PD: photodiode; PC: principal crystal; CC: compensation crystal; BS: beam splitter; RR: retro-reflector; TM: triangular mirror.}
	\label{fig:full_scheme} 
\end{figure}

Fig.~\ref{fig:full_scheme} shows a detailed, schematic representation of the experimental setup.
We generate heralded single photons via degenerate type-II Spontaneous Parametric Down Conversion (SPDC) occurring within a periodically-poled potassium titanyl phosphate (ppTKP) crystal.
The optical pump used for the SPDC generation is a CW diode laser at 405 nm set to a horizontal ($H$) polarization, with the orthogonally-polarized signal–idler photon pairs at 810 nm split in two different paths by a PBS.
The DM$_\text{P}$ dichroic mirror,
reflecting 405 nm photons while transmitting 810 nm ones, eliminates the pump light from the rest of the setup.
Idler photons are coupled into a multimode fiber and detected by a SPAD, heralding the presence of the corresponding signal photon in the other path.
A 3 nm bandwidth interference filter centered at 810 nm improves spectral purity, allowing to increase the photon coherence length to improve the further interference.
The heralded signal photons constituting our qubits are coupled to a single-mode fiber for spatial filtering and, subsequently, sent again in free space.
Then, the QWP$_\text{IN}$, the HWP$_\text{IN}$, and the PBS$_\text{IN}$ allow compensating fiber-induced polarization rotations, and a lens (L$_\text{IN}$) is used to collimate the photons spatial distribution along the entire DIQC.
The 50:50 BS$_\text{IN}$ sends half of the heralded photons to a monitor arm, where they are coupled to a multimode fiber and detected by the monitor detector SPAD$_\text{M}$, whose photon counts are used to keep track of the source intensity fluctuations during the data acquisition.
The remaining single-photon qubits are prepared in the arbitrary state $\ket{\psi}=\cos(\psi)\ket{H}+\sin(\psi)\ket{V}$ by the HWP$_\text{pr}$ and then addressed to the MZI.
Here, the MZI input PBS$_1$ realizes the coupling between the polarization and the two paths of the interferometer, acting as the ancillary DoFs:
\begin{equation}
\cos(\psi)|H\rangle+\sin(\psi)|V\rangle \;\mapsto\; \cos(\psi)|H\rangle|A\rangle+\sin(\psi)|V\rangle|B\rangle,
\label{eq:PBS_op}
\end{equation}
where \(|A\rangle\) and \(|B\rangle\) denote the spatial modes transmitted and reflected by PBS$_1$, respectively.
The HWPs (HWP$_{\text{A}_\text{IN}}$ and HWP$_{\text{B}_\text{IN}}$) at the entrance of each branch rotate the polarization to the QSUP-protected state $\ket\xi$, so that
\begin{equation}
\cos(\psi) \ket H \ket A +\sin(\psi) \ket V \ket B
\;\longmapsto\;
\ket\xi \otimes\big(\cos(\psi)\,|A\rangle+\sin(\psi)\,|B\rangle\big)\;.
\end{equation}
In each arm, the decoherence-inducing qubit-environment coupling is realized via birefringent \(\mathrm{CaCO}_3\) crystal pairs that correlate polarization with the photon transverse spatial DoF $x$, which plays the role of the environment (see Refs. \cite{Piacentini2017,Pia2021,reb2021,Virzi2024,reb2025} for details).
Between subsequent decoherence blocks, Glan-Laser polarizers mounted on rotation stages realize the QZE projections $\hat\Pi_\xi=\ketbra{\xi}{\xi}$.
At the DIQC output, the HWP$_{\text{A}_\text{OUT}}$ and HWP$_{\text{B}_\text{OUT}}$ in the A and B arms of the MZI, together with PBS$_2$, perform the reverse SWAP operation, restoring the qubit in its initial polarization state.
To coherently recombine the A and B modes, one has to match the two path lengths within the coherence length of the photons.
This is achieved by employing two motorized translation stages; one of the stages provides coarse path-length balancing, allowing to find the interference point, while the other one allows actively stabilizing and maintaining the phase over time.

\subsection*{DIQC realization in presence and absence of QSUP}

With our interferometric setup, we are able to implement the DIQC in both experimental configurations, i.e. with and without QSUP. \\
\textbf{QSUP-assisted DIQC.} The MZI arms A and B, acting as the ancillary modes, contain each an initial HWP$_{j_\text{IN}}$ (with $j=\text{A,B}$), a sequence of $k=0,...,N=4$ decoherence blocks interleaved with the polarizers realizing the QZE projections, and the final HWP$_{j_\text{OUT}}$.
The MZI input/output PBSs, together with the HWPs, allow realizing the qubit-ancilla SWAP operations.
At Bob’s station, a polarization analysis setup composed of HWP$_\text T$, QWP$_\text T$ and PBS$_\text T$ is used to perform single-qubit quantum state tomography.
After the PBS$_\text T$, photons are coupled to a multimode fiber and detected by the SPAD$_\text T$.
A 3 nm-bandwidth interference filter centered at 810 nm suppresses background light before fiber coupling.\\
\textbf{DIQC with no QSUP.} By preparing $V$-polarized photons, the photons travel only in the arm B of the MZI, reproducing the dynamics of a traditional DIQC.
Then, the initial HWP$_{\text{B}_\text{IN}}$ is used to prepare the input polarization state $\ket{\psi}$. The same sequence of $k$ decoherence blocks implemented in the previous case follows. Finally, HWP$_{\text{B}_\text{OUT}}$, an additional QWP, PBS$_2$ and Bob's detector are exploited to perform the  quantum state tomography.

\subsection*{Data acquisition and analysis}

Polarization quantum state tomography is performed on the output qubit for each input state $\ket{\psi}\in\{\ket{20^\circ},\ket{45^\circ},\ket{60^\circ}\}$ and each QSUP  state $\ket{\xi}\in\{\ket{20^\circ},\ket{45^\circ},\ket{60^\circ}\}$, for $k=0,\dots,4$ decoherence blocks in the DIQC.
The quantum tomography apparatus projects the single-photon qubit onto the six polarization states $\{\ket H,\ket V,\ket D=\frac{\ket H + \ket V}{\sqrt2},\ket A=\frac{\ket H - \ket V}{\sqrt2},\ket L=\frac{\ket H + i\ket V}{\sqrt2},\ket R=\frac{\ket H - i\ket V}{\sqrt2}\}$.
For each projection, $N_T^{(k)}=30$ measurements are collected, with $1\,\text{s}$ integration time each, registering also the monitor photon counts.
For each $i$ value, and each number of decoherence blocks $k$, the density matrix $\rho_i^{(k)}$ of the qubit state exiting the DIQC is reconstructed using the output coincidences recorded for all measurement basis, by exploiting a maximum-likelihood method \cite{ParisRehacek2004}.
For each configuration, the mean Fidelity $\mathcal F^{(k)}$ with respect to the input state $\ket{\psi}$, and its associated uncertainty, are obtained as:
\begin{equation}
\mathcal F^{(k)}=\frac{1}{N_T^{(k)}}\sum_{i=1}^{N_T^{(k)}} \mathcal F_i^{(k)}\;,\qquad
\mathcal F_i^{(k)}=\mathrm{Tr}\big[\rho_i^{(k)}\,|\psi\rangle\!\langle\psi|\big]\;,
\label{eq:fid_ave}
\end{equation}
\begin{equation}
\delta \mathcal F^{(k)}=\sqrt{\frac{1}{N_T^{(k)}(N_T^{(k)}-1)}\sum_{i=1}^{N_T^{(k)}}\big(\mathcal F_i^{(k)}- \mathcal F^{(k)}\big)^2}\;
\label{eq:fid_unc}
\end{equation}
Conversely, the average Purity $\mathcal P^{(k)}$ of the qubit output state, and its associated statistical uncertainty $\delta\mathcal P^{(k)}$, are expressed as:
\begin{equation}
\mathcal P^{(k)}=\frac{1}{N_T^{(k)}}\sum_{i=1}^{N_T^{(k)}} \mathcal P_i^{(k)}\;,\qquad
\mathcal P_i^{(k)}=\mathrm{Tr}\big[(\rho_i^{(k)})^2\big]\;,
\label{eq:pur_ave}
\end{equation}
\begin{equation}
\delta \mathcal P^{(k)}=\sqrt{\frac{1}{N_T^{(k)}(N_T^{(k)}-1)}\sum_{i=1}^{N_T}\big(\mathcal P_i^{(k)} - \mathcal P^{(k)}\big)^2}\;.
\end{equation}
Finally, the experimental survival probability $p_\text{sur} (k)$, shown in Fig. \ref{fig:data}(e), is evaluated as
\begin{equation}
p_\text{sur}(k) = (1-\mathcal{L}^{(k)})\;,
\end{equation}
with the loss factor \(\mathcal{L}^{(k)}\) estimated from the heralded photon count rates:
\begin{equation}
\mathcal{L}^{(k)}=\frac{L_k}{L_0}\;,\qquad
L_k=\frac{C_{H}^{(k)}}{M_{H}^{(k)}}+\frac{C_{V}^{(k)}}{M_{V}^{(k)}}+\frac{C_{D}^{(k)}}{M_{D}^{(k)}}+\frac{C_{A}^{(k)}}{M_{A}^{(k)}}+\frac{C_{R}^{(k)}}{M_{R}^{(k)}}+\frac{C_{L}^{(k)}}{M_{L}^{(k)}}\;.
\end{equation}
Here, $k$ is the number of decoherence blocks present in the DIQC, $C_{\zeta}^{(k)}$ is the number of photon counts at the DIQC output (averaged over the 30 repetitions for each tomographic projection setting $\zeta$), and $M^{(k)}$ is the corresponding number of photon counts occurring in the monitor arm (to compensate for eventual drifts in the heralded photon source output power).
The reference maximum transmittance $L_0$ is obtained by removing all decoherence blocks in the DIQC.
Reported survival probability values include both passive optical losses and projection-induced losses due to the finite size of the steps.


\clearpage 




\end{document}